\documentclass[12pt]{article}

\RequirePackage{etex}

\usepackage[margin=1in]{geometry}  	
\usepackage{amsmath,amsfonts,amssymb,amsthm}
\usepackage{booktabs}
\usepackage{multirow}
\usepackage[onehalfspacing]{setspace} 
\renewcommand{\arraystretch}{0.6} 

\usepackage{tcolorbox}
\usepackage{tabularx}
\usepackage{threeparttable}

\usepackage{float}
\usepackage{caption}
\usepackage{graphicx}
\usepackage{adjustbox}
\usepackage{lscape}

\usepackage{xcolor}

\usepackage{relsize}

\usepackage{bbm}
\usepackage{subfig}

\usepackage{titlesec}

\usepackage{hyperref}

\usepackage{authblk}

\usepackage{amsthm}

\newtheorem*{proposition*}{Proposition}

\usepackage{enumitem}

\newlist{myenum}{enumerate}{1}
\setlist[myenum,1]{label=\Alph{itemalpha}\arabic{itemnum}, ref=\Alph{itemalpha}\arabic{itemnum}}
\newcounter{itemalpha}
\newcounter{itemnum}

\usepackage{xpatch}
\makeatletter
\xpatchcmd{\proof}{\@addpunct{.}}{\@addpunct{:}}{}{}
\makeatother

\makeatletter
\newcommand{\vast}{\bBigg@{3}}
\newcommand{\Vast}{\bBigg@{4}}
\makeatother

\makeatletter
\newcommand*{\indep}{%
  \mathbin{%
    \mathpalette{\@indep}{}%
  }%
}
\newcommand*{\nindep}{%
  \mathbin{
    \mathpalette{\@indep}{\not}
  }%
}
\newcommand*{\@indep}[2]{%
  \sbox0{$#1\perp\m@th$}
  \sbox2{$#1=$}
  \sbox4{$#1\vcenter{}$}
  \rlap{\copy0}
  \dimen@=\dimexpr\ht2-\ht4-.2pt\relax
  \kern\dimen@
  {#2}%
  \kern\dimen@
  \copy0 
} 
\makeatother

\DeclareMathOperator{\E}{\textnormal{\mbox{E}}}

\usepackage[utf8]{inputenc}
\DeclareUnicodeCharacter{200E}{}

\usepackage{setspace}
\usepackage{cite}

\usepackage{xr-hyper}

\usepackage{xcolor}
\definecolor{forestgreen}{RGB}{34,139,34}
\usepackage{hyperref}
\hypersetup{
    colorlinks=true,
    linkcolor=magenta,
    filecolor=magenta, 
    citecolor=forestgreen,      
    urlcolor=blue
}

\makeatletter
\renewcommand{\boxed}[1]{\text{\fboxsep=.2em\fbox{\m@th$\displaystyle#1$}}}
\makeatother

\makeatletter
\newcommand*{\addFileDependency}[1]{
  \typeout{(#1)}
  \@addtofilelist{#1}
  \IfFileExists{#1}{}{\typeout{No file #1.}}
}
\makeatother

\newcommand*{\myexternaldocument}[1]{%
    \externaldocument{#1}%
    \addFileDependency{#1.tex}%
    \addFileDependency{#1.aux}%
}

\myexternaldocument{appendix}

\usepackage[stable]{footmisc}

\usepackage{rotating}

\makeatletter
\def\@hangfrom#1{\setbox\@tempboxa\hbox{{#1}}%
      \hangindent 0pt
      \noindent\box\@tempboxa}
\makeatother

\makeatletter
\def\@seccntformat#1{\@ifundefined{#1@cntformat}%
   {\csname the#1\endcsname\quad}  
   {\csname #1@cntformat\endcsname}
}
\let\oldappendix\appendix 
\renewcommand\appendix{%
    \oldappendix
    \newcommand{\section@cntformat}{\appendixname~\thesection\quad}
}
\makeatother

\usepackage[absolute,showboxes]{textpos}

\TPMargin{5pt}

\newcommand{\copyrightstatement}{
    \begin{textblock}{0.84}(0.08,0.93)    
         \noindent
         \footnotesize
         This DRAFT presents WORK IN PROGRESS. Please send comments to \href{mailto:idahabreh@hsph.harvard.edu}{idahabreh@hsph.harvard.edu}.
    \end{textblock}
}

\usepackage{datetime}
\def\paperversionmajor{3}
\def\paperversionminor{0}

\usepackage{comment}
\begin{document}
\pagenumbering{gobble}

\title{Comparing randomized trials: a formal approach}

\author[1,2]{Yu-Han Chiu}
\author[3,4]{Issa J. Dahabreh}

\affil[1]{Department of Epidemiology, Brown University School of Public Health, Providence, RI}
\affil[2]{Center for Gerontology and Health Care Research, Brown University School of Public Health, Providence, RI}
\affil[3]{CAUSALab, Department of Epidemiology, and Department of Biostatistics, Harvard T.H. Chan School of Public Health, Boston, MA}
\affil[4]{Richard A. and Susan F. Smith Institute Research for Evidence, Innovation, and Policy, Beth Israel Deaconess Medical Center, Harvard Medical School, Boston, MA}

\date{}

\maketitle
\copyrightstatement


\vspace*{0.3in}

\begin{abstract}
\noindent
\linespread{1.4}\selectfont
Disagreements between randomized trials that address the same causal question pose challenges for the interpretation and synthesis of evidence. We sketch a causal framework for comparing the results of randomized trials that explicitly accommodates the possibility of trial engagement effects. Using counterfactual outcomes defined under joint interventions on treatment assignment and trial-related actions, we compare trial-native treatment effects with the effects transported from one trial to the population of another in a common target population. We show that standard conditions for transporting treatment effects across trial populations are not sufficient to guarantee agreement between transported and trial-native effects. Agreement additionally requires conditions governing how trial-related actions affect outcome across studies—for example, absence of trial engagement effects, equivalence of such effects across trials, or lack of interaction between trial engagement and treatment assignment. Our framework clarifies potential sources of disagreement across trials and extends existing approaches to study comparison to settings in which trial engagement effects may be present.

\vspace{0.2in}
\noindent
\textbf{Keywords:} randomized trials; transportability; generalizability; combining information; trial engagement effects.

\end{abstract}

\clearpage
\section*{}
\pagenumbering{arabic}

\noindent

Formal comparisons across studies—including trial comparisons, benchmarking analyses, and experimental replication efforts \cite{dahabreh2020benchmarking, wong2022design}—support large-scale empirical evaluations of concordance and discordance between studies. Such comparisons are central to the interpretation and synthesis of evidence because studies addressing ostensibly the same intervention may differ in populations, health systems, time periods, or trial implementation protocols.

Prior work has proposed causal frameworks for benchmarking observational analyses against randomized trials \cite{dahabreh2020benchmarking}. These frameworks describe conditions under which agreement between observational and randomized analyses is expected, typically assuming that trial participation affects outcomes only through treatment assignment. In practice, however,  trial participation may influence outcomes not mediated through treatment assignment, such as through additional monitoring, clinical counseling, or ancillary interventions \cite{rothwell2005external,dekkers2010assess,dodd2021platform}. We refer to such effects as trial engagement effects \cite{dahabreh2019identification, dahabreh2022generalizing}.

Here, we sketch a causal framework for comparing randomized trials that explicitly accommodates the possibility of trial engagement effects \cite{ung2025generalizing}.
The framework clarifies that agreement between trials depends not only on identification conditions analogous to those used in benchmarking observational analyses against trials \cite{dahabreh2020benchmarking}, but also on additional conditions governing how trial-related actions influence outcomes across studies. More broadly, the ideas developed here apply not only to comparisons between randomized trials, but also to benchmarking and experimental replication efforts.


\section*{Data structure and sampling} 

Suppose we have data from two separately conducted randomized trials, each enrolling participants from distinct and possibly different populations. We use $S \in \{0,1\}$ to denote trial membership, $A \in \{0,1\}$ to denote binary treatment assignment, $Y$ to denote the outcome (e.g., continuous, binary, or count) measured at the end of follow-up, $X$ the vector of measured baseline covariates, and $U$ the vector of unmeasured covariates. We model the data from the two trials as independently sampled from distinct underlying populations, with unknown and potentially unequal sampling fractions across populations (i.e., a non-nested design \cite{dahabreh2019towards,dahabreh2021study}). 

When referring generically to the two trials, we use $s$ to indicate one trial and $1-s$ to index the other, for $s \in \{0,1\}$. When one trial is used as the source of information about the treatment--outcome association, we refer to it as the index or source trial and to the other as the target trial. The roles of the trials can be reversed, provided the required conditions hold in both directions. To simplify the exposition, we assume no loss to follow-up and no measurement error (in the sense that we consider each trial as able to measure the variables it chooses to measure).

\section*{Causal estimands for trial comparisons} 
We use $Y^{s,a}$ to denote the potential (counterfactual) outcome under a joint intervention that sets treatment assignment to $A=a$ and sets the outcome-relevant trial actions and non-randomized interventions to those of trial $S = s$ \cite{balzer2017all, dahabreh2019identification}. Considering such an intervention may be appropriate because outcomes may depend on aspects of the trial other than the randomized treatment itself \cite{dahabreh2020extending}. We refer to the effects of trial-related actions and interventions that are not mediated by the randomized treatment as \emph{trial engagement effects} \cite{dahabreh2019identification, ung2025generalizing}. Trial engagement effects may arise, for example, from additional monitoring, clinical counseling, or ancillary interventions provided as part of study participation. We use $Y^a$ to denote the potential outcome under intervention only to set treatment $A$ to $a$; these potential outcomes are useful, for example, when trial engagement effects are absent. 

One family of causal estimands of interest is the average effect of assignment to treatment versus control under the trial-related actions and interventions of a given trial $s \in \{0,1\}$, among the populations underlying either trial: 

\begin{align*}
\E\!\left[Y^{s,a=1} - Y^{s,a=0} \mid S = s\right] 
\quad \text{or} \quad 
\E\!\left[Y^{1-s,a=1} - Y^{1-s,a=0} \mid S = s\right].
\end{align*}

The corresponding covariate-conditional estimands are
\begin{align*}
\E\!\left[Y^{s,a=1} - Y^{s,a=0} \mid X, S = s\right] 
\quad \text{or} \quad 
\E\!\left[Y^{1-s,a=1} - Y^{1-s,a=0} \mid X, S = s\right].
\end{align*}


These estimands define four causal quantities corresponding to the effect of treatment under the trial-related actions of each trial, evaluated in either trial population (Table~\ref{tab:causal}). For $s \in \{0,1\}$, the diagonal elements correspond to causal effects under the trial-related actions and interventions of trial $s$ in its own population $s$, whereas the off-diagonal elements correspond to causal effects under the trial-related actions and interventions of the other trial $1-s$, evaluated in that same population $s$. Formal comparisons between trials therefore require additional conditions under which causal effects defined under different trial-related conditions are equal within the same population.

Throughout, all estimands are conditional on trial membership because we view the two trials as samples from different underlying populations. In principle, one could also define estimands in a mixture of the two populations, but the observed data typically do not provide information about the sampling fractions from the underlying populations into the two trials, and therefore it is unlikely that a mixture of the trial populations corresponds to a real-world population \cite{dahabreh2021study, dahabreh2019towards}.

\section*{Identifiability conditions for comparing randomized trials}\label{sec:conditions}
In this section, we formalize a set of sufficient conditions for identifying and comparing causal effects across randomized trials.


Broadly, they are of two kinds: The first kind ensures that each trial can identify the causal effects of interest in its underlying population and can be validly transported across trials. These considerations parallel those used in frameworks for benchmarking observational analyses against trials \cite{dahabreh2020benchmarking, dahabreh2022randomized}. The second ensures that causal effects under different trial settings are comparable in a common target population.  Together, these conditions support formal cross-trial comparisons.

\subsection*{Conditions for identification of causal effects within trials and for transporting causal effects across trials}

\paragraph{A1. Consistency of potential outcomes with respect to interventions on trial participation $S$ and treatment assignment $A$.} If an individual participates in trial $s$ and is assigned treatment $a$, then the observed outcome equals the potential outcome under the joint intervention that sets treatment assignment to $A=a$ and sets trial-related actions to those of trial $S=s$. That is, if $A=a$ and $S=s$, then
\[
Y = Y^{s,a}, \qquad a\in\{0,1\} \text{ and} \; s\in\{0,1\}.
\]

\paragraph{A2. Conditional exchangeability for treatment assignment $A$ within trial.} Within each trial, conditional on covariates, the potential outcome under the joint intervention that sets trial participation to $s$ and treatment assignment to $a$ is independent of observed treatment assignment. That is, for $s \in \{0,1\}$,
\[
Y^{s,a} \indep A \mid X, S=s.
\]


\paragraph{A3. Positivity of treatment assignment $A$ within each trial.}
Within each trial, for every covariate value $x$ observed in that trial, each treatment level has a positive probability of assignment. That is, 
\[
\Pr(A=a \mid X=x, S=s) > 0
\]
for all $x$ such that $f_{X,S}(x,s) \neq 0$ and $a \in \{0,1\}$.

Conditions A1–A3 are standard conditions for identification of causal effects within each trial, $\E\!\left[Y^{s,a=1} - Y^{s,a=0} \mid S = s\right]$, and are expected to hold by design in a marginally or conditionally randomized trial. Now we describe
additional conditions required for extending inference across trials, $\E\!\left[Y^{1-s,a=1} - Y^{1-s,a=0} \mid S = s\right]$.

\paragraph{A4. No effect modification by trial membership, conditional on covariates $X$.}
For every covariate value $x$ with positive density in the target trial, the effect of treatment assignment under the trial-related conditions of trial $s$ is the same in the source and target trial populations, conditional on $X$. That is, for all $x$ such that $f_{X,S}(x,s) \neq 0$ and $s \in \{0,1\}$,
\[
\E[Y^{s,a=1} - Y^{s,a=0} \mid X=x, S=s]
=
\E[Y^{s,a=1} - Y^{s,a=0} \mid X=x, S=1-s].
\]

\paragraph{A5. Positivity of trial membership.}
For every covariate value $x$ with positive density in the target trial population $s$, there must be a nonzero probability of observing individuals with those covariate values in the source trial $1-s$:
\[
\Pr(S=1-s \mid X=x) > 0
\]
for all $x$ such that $f_{X,S}(x,s) \neq 0$.

\subsection*{Conditions for comparability of causal effects across trials}
To compare causal effects defined under different trial settings (i.e., diagonal and off-diagonal elements within the same row of Table~\ref{tab:causal}), additional conditions about trial engagement effects are required. Assumptions B1--B3 below describe progressively weaker sufficient conditions under which causal effects defined under different trial-related conditions coincide within the same population. For interpretability, these conditions are stated in terms of individual-level potential outcomes. For identification and comparison, it is sufficient that the corresponding equalities hold at the level of conditional expectations.

\paragraph{B1. Absence of trial engagement effects.}
When trial engagement effects are absent, such that $Y^{s,a}=Y^{a}$, potential outcomes do not depend on trial membership. This assumption may be plausible, for example, in pragmatic trials where the experimental process does not modify aspects of care other than the randomized treatments. In this case,
\[
\E[Y^{s,a}\mid X=x,S=s]
=
\E[Y^{1-s,a}\mid X=x,S=s]
=
\E[Y^{a}\mid X=x,S=s],
\]
and therefore
\begin{equation*}
\E[Y^{s,a=1} - Y^{s,a=0} \mid X, S=s]
=\E[Y^{1-s,a=1} - Y^{1-s,a=0} \mid X, S=s] =\E[Y^{a=1} - Y^{a=0} \mid X, S=s].
\end{equation*}

\paragraph{B2. Trial engagement effects are present but the same across trials.} When trial engagement effects are present, but the same across trials, such that for all $a$, $Y^{s,a} = Y^{1-s,a}$,
the effects of trial-related actions and non-randomized interventions affect outcomes similarly across trial settings. This assumption may be plausible when trials are conducted in a similar manner—for example, as part of a coordinated research program with comparable protocols, monitoring, and co-interventions beyond the randomized treatment assignment. Under this condition, for every covariate value $x$ such that $f_{X,S}(x,s) \neq 0$,
\[
\E[Y^{s,a}\mid X=x,S=s]
=
\E[Y^{1-s,a}\mid X=x,S=s],
\]
and therefore,
\begin{equation*}\label{eq:same_engagement_appendix}
\E[Y^{s,a=1}-Y^{s,a=0}\mid X=x, S=s]
=
\E[Y^{1-s,a=1}-Y^{1-s,a=0}\mid X=x, S=s].
\end{equation*}

\paragraph{B3. Trial engagement effects differ across trials, but do not interact with treatment assignment.}
When trial engagement effects differ across trials—for example, due to differences in clinical practices or co-interventions across settings or time periods—but there is no interaction between trial-related conditions and treatment assignment on the additive scale, the effect of intervening to set treatment $A$ to $a=1$ versus $a=0$ is the same under the trial-related actions of either trial. That is, for each individual,
$Y^{s,a=1} - Y^{s,a=0} = Y^{1-s,a=1} - Y^{1-s,a=0}.$
This assumption may be plausible when differences in clinical practices or co-interventions across trial settings affect outcomes similarly under both treatment levels. Under this condition, for all covariate values $x$ such that $f_{X,S}(x,s) \neq 0$,
\begin{equation*}\label{eq:no_interaction_conditional_appendix}
\E[Y^{s,a=1}-Y^{s,a=0}\mid X=x, S=s]
=
\E[Y^{1-s,a=1}-Y^{1-s,a=0}\mid X=x, S=s].
\end{equation*}
This assumption is weaker than conditions B1 and B2 because it requires equality only of treatment effects, not of the underlying potential outcomes. 

\section*{Comparing the results of two trials}
We now describe how the conditions together support statistical comparisons of effect estimates across trials. In the main text, we present results using nested expectation expressions (g-formula type expression \cite{robins1986new}); equivalent weighting representations and proofs are provided in the Appendix.

We first consider identification of trial-native effects within each trial, then identification of transported effects across trial populations, and finally conditions under which transported and trial-native effects coincide within the same population.

\paragraph{Identification in each trial.}
Under the conditions A1--A3 described above, it is well-known that, for each trial $s \in \{0,1\}$, the conditional average effect of treatment assignment under the outcome-relevant actions and interventions of trial $s$, among participants in trial $s$, is identified by
\begin{equation}\label{eq:direct_conditional}
\E[Y^{s,a=1} - Y^{s,a=0} \mid X, S=s]
=
\E[Y \mid X, S=s, A=1] - \E[Y \mid X, S=s, A=0].
\end{equation}

The right-hand side of the expression above is the difference in the conditional mean outcome between the two treatment groups within trial $s$, among participants with the same covariate values $X$. Identification of this conditional effect therefore requires only observed data from trial $s$.

By the law of total expectation, the corresponding average effect of assignment in trial $s$ is identified by
\begin{equation}\label{eq:direct_marginal}
\begin{split}
\E\!\left[Y^{s,a=1} - Y^{s,a=0} \mid S=s\right] =
\E\!\left[\E[Y \mid X, S=s, A=1] - \E[Y \mid X, S=s, A=0] \mid S=s\right].
\end{split}
\end{equation}
That is, the marginal effect in trial $s$ is obtained by averaging the conditional effect over the covariate distribution in trial $s$. The conditional outcome expectations are estimated using only data from trial $s$, and the marginalization also uses only the covariate distribution in that trial.

\paragraph{Identification by transporting effects across trials.}
We next consider transporting effects across trials. For the population underlying each trials $s \in \{0,1\}$, a different causal effect is identified by transporting information from the other trial $s'$, where $s' = 1 - s$, to the population of trial $s$. This effect involves the outcome-relevant actions and interventions of trial $s'$ together with treatment assignment, but applied to the population of trial $s$. The identification of this effect requires additional conditions, namely A4--A5, in addition to those needed for identification within each trial.

Under the conditions A1--A5, for each trial $s \in \{0,1\}$, the conditional effect of the joint intervention that implements the outcome-relevant actions and interventions of trial $s' = 1 -s$ together with treatment assignment, in the population of trial $s$, is identified by the observed-data conditional expectation difference in trial $s'$:
\begin{equation}\label{eq:transport_conditional}
\begin{split}
\E[Y^{s',a=1} - Y^{s',a=0} \mid X, S=s] &=
\E[Y^{1-s,a=1} - Y^{1-s,a=0} \mid X, S=s]  \\
&= \E[Y \mid X, S=1-s, A=1] - \E[Y \mid X, S=1-s, A=0].
\end{split}
\end{equation}
The observed-data functional on the right-hand side is the difference in the conditional mean outcome between the two treatment groups in trial $s'  = 1-s$, among participants with the same covariate values $X$. Thus, the transported conditional effect for the population of trial $s$ is identified using observed outcome information from trial $1-s$.

By the law of total expectation, the corresponding marginal effect in the population of trial $s \in \{0,1\}$, under the outcome-relevant actions and interventions of trial $s' = 1- s$, is identified by
\begin{equation}\label{eq:transport_marginal_other}
\begin{split}
\E\!\left[Y^{s',a=1} - Y^{s',a=0} \mid S=s\right] &= 
\E\!\left[Y^{1-s,a=1} - Y^{1-s,a=0} \mid S=s\right] \\
&= \E\!\left[\E[Y \mid X, S=1-s, A=1] - \E[Y \mid X, S=1-s, A=0] \mid S=s\right].
\end{split}
\end{equation}
That is, the marginal transported effect is obtained by averaging the conditional effect over the covariate distribution in trial $s$. The conditional outcome expectations are obtained from trial $s' = 1- s$, whereas the marginalization uses the covariate distribution in trial $s$.

At this stage, the transported estimand may not equal the trial-native estimand in trial $s$ because trial engagement effects may differ between the two trials. 
That is, for each $a$, $Y^{s,a}$ may not equal $Y^{1-s,a}$ leading to differences between $Y^{s,a=1}-Y^{s,a=0}$ and $Y^{1-s,a=1}-Y^{1-s,a=0}$, as well as differences in the (conditional or marginal) expectations of these contrasts in the population underlying either trial.

\paragraph{Comparing effects in the same trial population.}
Equality between trial-native and transported effects within the same population generally requires additional assumptions beyond those described above. Conditions B1--B3 describe distinct sufficient conditions under which such equality holds. For example, under condition B3  (no additive interaction between trial engagement and treatment assignment), the effect of treatment assignment is the same under the trial-related conditions of either trial \cite{ung2025generalizing}. That is,
\begin{equation*}\label{eq:no_interaction_conditional}
\E[Y^{s,a=1} - Y^{s,a=0} \mid X, S=s]
=
\E[Y^{1-s,a=1} - Y^{1-s,a=0} \mid X, S=s].
\end{equation*}
Under this additional condition, the transported conditional effect and the trial-native conditional effect coincide in the population of trial $s$. Therefore, the conditional effect under the outcome-relevant actions and interventions of trial $s$, in the population of trial $s$, can also be identified by the observed-data conditional expectation difference in trial $1-s$:
\begin{equation}\label{eq:transport_conditional_same}
\E[Y^{s,a=1} - Y^{s,a=0} \mid X, S=s]
=
\E[Y \mid X, S=1-s, A=1] - \E[Y \mid X, S=1-s, A=0].
\end{equation}

By the law of total expectation,
\begin{equation}\label{eq:transport_marginal}
\begin{split}
\E[Y^{s,a=1} - Y^{s,a=0} \mid S=s] =
\E\!\left[\E[Y \mid X, S=1-s, A=1] - \E[Y \mid X, S=1-s, A=0] \mid S=s\right].
\end{split}
\end{equation}

That is, the same causal estimand---the marginal effect of assignment to treatment versus control under the outcome-relevant actions and interventions of trial $s$, in the population of trial $s$---can be identified either using data from trial $s$ itself or by combining the conditional outcome expectations from trial $1-s$ with the covariate distribution in trial $s$. Accordingly, under these conditions, the two identifying functionals correspond to the same causal quantity.

The same identification results follow under conditions B1 and B2. Notably, the observed-data identifying functionals under the weaker no-interaction condition (B3) are algebraically identical to those obtained under the stronger conditions B1 and B2, despite relying on different causal assumptions. Prior benchmarking frameworks typically focus on settings corresponding to condition B1.

\paragraph{Using the data to compare.}
The identifying expressions above imply null hypotheses that can be examined using observed data. At the conditional level, for each $x$ in the region of covariate overlap between the trials, the null hypothesis of agreement is
\begin{equation}\label{eq:null_conditional}
\begin{split}
&\E[Y \mid X, S=0, A=1] - \E[Y \mid X, S=0, A=0] \\
&\qquad = \E[Y \mid X = x, S=1, A=1] - \E[Y \mid X=x, S=1, A=0].
\end{split}
\end{equation}
That is, the conditional mean difference comparing treatment groups is the same in the two trials at the same covariate values. Under the conditions described above, this observed-data null corresponds to equality of the conditional causal effects in the same trial population.

More commonly, however, the comparison is made at the marginal level. For a given trial $s$, one may compare the effect identified directly in that trial with the effect obtained by transporting information from trial $s'=1-s$ to trial $s$. The corresponding null hypotheses are, for $s \in \{0,1\}$,
\begin{equation}\label{eq:null_marginal}
\begin{split}
&\E\!\left[\E[Y \mid X, S=s, A=1] - \E[Y \mid X, S=s, A=0] \mid S=s\right] \\
&\qquad =
\E\!\left[\E[Y \mid X, S=1-s, A=1] - \E[Y \mid X, S=1-s, A=0] \mid S=s\right].
\end{split}
\end{equation}
That is, the marginal effect of assignment to treatment versus control under the outcome-relevant actions and interventions of trial $s$, in the population of trial $s$, obtained using data from trial $s$ itself, agrees with the corresponding effect obtained by combining the conditional outcome expectations from trial $1-s$ with the covariate distribution of trial $s$. 


In principle, one could examine the conditional-on-covariates null hypothesis in equation~\eqref{eq:null_conditional} as agreement between trials is most naturally defined within levels of baseline covariates. In practice, however, conditional comparisons are often difficult because they require estimation and inference for covariate-specific effects, which may be unstable when covariates are high-dimensional or overlap is limited. Arguably, that is the reason why formal comparisons between trials often focus on the marginal null hypothesis in equation~\eqref{eq:null_marginal}. 


\section*{Conclusion}

We sketched a causal framework for comparing results across randomized trials and interrogating sources of disagreement between them. By defining estimands under joint interventions on treatment assignment and trial participation, cross-trial comparisons can be understood as comparisons between trial-native effects and transported effects within a common population. The framework clarifies that transportability conditions alone does not, in general, ensure agreement between trials: additional conditions regarding trial engagement effects are required for transported and trial-native effects to coincide.

One potential application of this framework, with appropriate extensions to address practical challenges such as non-adherence, missing data, and censoring, is the systematic evaluation of similarly designed trials addressing the same clinical question. Such analyses may help interrogate disagreements between studies and may inform the design, interpretation, and synthesis of future evidence.

The present framework focuses on the causal conditions under which agreement between studies is expected. The interpretation of empirical comparisons may also depend on how such comparisons are selected; related statistical considerations are discussed elsewhere.

\clearpage
\begin{table}[htbp]
\centering
\begin{threeparttable}
\renewcommand{\arraystretch}{1.3}
\setlength{\tabcolsep}{6pt}
\caption{Causal quantities of interest in cross-trial comparison.}
\label{tab:causal}

\begin{tabularx}{\textwidth}{
>{\raggedright\arraybackslash\hsize=0.45\hsize}X
>{\raggedright\arraybackslash\hsize=1.15\hsize}X
>{\raggedright\arraybackslash\hsize=1.15\hsize}X
}
\hline
& \textbf{Joint interventions: $(s=1,a=1)$ vs $(s=1,a=0)$}
& \textbf{Joint interventions: $(s=0,a=1)$ vs $(s=0,a=0)$} \\
\hline

\textbf{Conditional on $S=1$} 
& $\E\!\left[Y^{s=1,a=1}-Y^{s=1,a=0}\mid X, S=1\right]$
& $\E\!\left[Y^{s=0,a=1}-Y^{s=0,a=0}\mid X, S=1\right]$ \\[6pt]

\textbf{Conditional on $S=0$} 
& $\E\!\left[Y^{s=1,a=1}-Y^{s=1,a=0}\mid X, S=0\right]$
& $\E\!\left[Y^{s=0,a=1}-Y^{s=0,a=0}\mid X, S=0\right]$ \\

\hline
\end{tabularx}
\begin{tablenotes}
\footnotesize
\item The diagonal elements correspond to trial-specific effects identified within each trial, whereas the off-diagonal elements correspond to effects obtained by transporting results across trials.
\end{tablenotes}

\end{threeparttable}
\end{table}

\section*{Acknowledgments}


This work was supported in part by Patient-Centered Outcomes Research Institute (PCORI) award ME-2021C2-22365, National Library of Medicine award R01LM013616, National Heart, Lung, and Blood Institute award R01HL136708, and NIGMS R35GM154888.  \\
The content of this paper is solely the responsibility of the authors and does not necessarily represent the official views of PCORI, PCORI's Board of Governors, the PCORI Methodology Committee, or the National Institutes of Health.

\clearpage
\bibliographystyle{unsrt}
\bibliography{bibliography}{}


\ddmmyyyydate 
\newtimeformat{24h60m60s}{\twodigit{\THEHOUR}.\twodigit{\THEMINUTE}.32}
\settimeformat{24h60m60s}
\begin{center}
\vspace{\fill}\ \newline
\textcolor{black}{{\tiny $ $comparing\_trials, $ $ }
{\tiny $ $Date: \today~~ \currenttime $ $ }
{\tiny $ $Revision: \paperversionmajor.\paperversionminor $ $ }}
\end{center}

\end{document}